\documentclass[sigconf]{acmart}

\usepackage{multirow}
\usepackage{todonotes}
\usepackage{xcolor}
\usepackage{soul}

\AtBeginDocument{%
  \providecommand\BibTeX{{%
    \normalfont B\kern-0.5em{\scshape i\kern-0.25em b}\kern-0.8em\TeX}}}

\setcopyright{rightsretained} 
\copyrightyear{2019}
\acmYear{2019}
\acmDOI{10.1145/xxxxxxx.xxxxxxx}

\acmConference[EARS '19]{EARS '19: The 2nd International Workshop on Explainable Recommendation and Search (EARS 2019)}{July 25, 2019}{Paris, France}
\acmPrice{15.00}
\acmISBN{978-x-xxxx-xxxx-x/YY/MM}

\begin{document}

%%
%% A quick hack for notes and comments belonging to the underlined
%% text
\newcommand{\note}[2]
{
\setulcolor{red}
\ul{#1}\todo[size=\tiny]{#2}
}

\title{Effects of Foraging in Personalized Content-based Image Recommendation}

\author{Amit Kumar Jaiswal}
%\email{webmaster@marysville-ohio.com}
\affiliation{%
  \institution{University of Bedfordshire}
  \streetaddress{P.O. Box 1212}
  \city{Luton}
  %\state{Bedfordshire}
  \country{United Kingdom}}
\email{amitkumar.jaiswal@beds.ac.uk}

\author{Haiming Liu}
\affiliation{%
  \institution{University of Bedfordshire}
%  \streetaddress{1 Th{\o}rv{\"a}ld Circle}
  \city{Luton}
  \country{United Kingdom}}
\email{haiming.liu@beds.ac.uk}

\author{Ingo Frommholz}
\affiliation{%
  \institution{University of Bedfordshire}
  \city{Luton}
  \country{United Kingdom}}
\email{ingo.frommholz@beds.ac.uk}

\renewcommand{\shortauthors}{Jaiswal et al.}

\begin{abstract}
A major challenge of recommender systems is to help users locating interesting items. Personalized recommender systems have become very popular as they attempt to predetermine the needs of users and provide them with recommendations to personalize their navigation. However, few studies have addressed the question of what drives the users' attention to specific content within the collection and what influences the selection of interesting items. To this end, we employ the lens of Information Foraging Theory (IFT) to image recommendation to demonstrate how the user could utilize visual bookmarks to locate interesting images. We also found that the recommended image collection with visual bookmarks (cues) leads to a stronger scent. In this paper, we investigate a personalized content-based image recommendation system to understand what expropriates user attention by reinforcing visual attention based on IFT. Our evaluation is based on the Pinterest image collection

\end{abstract}
\begin{CCSXML}
<ccs2012>
<concept>
<concept_id>10002951</concept_id>
<concept_desc>Information systems</concept_desc>
<concept_significance>500</concept_significance>
</concept>
<concept>
<concept_id>10002951.10003317.10003331.10003271</concept_id>
<concept_desc>Information systems~Personalization</concept_desc>
<concept_significance>500</concept_significance>
</concept>
<concept>
<concept_id>10002951.10003317.10003347.10003350</concept_id>
<concept_desc>Information systems~Recommender systems</concept_desc>
<concept_significance>500</concept_significance>
</concept>
<concept>
<concept_id>10003120.10003121.10003124.10010868</concept_id>
<concept_desc>Human-centered computing~Web-based interaction</concept_desc>
<concept_significance>300</concept_significance>
</concept>
</ccs2012>
\end{CCSXML}

\ccsdesc[500]{Information systems}
\ccsdesc[500]{Information systems~Personalization}
\ccsdesc[500]{Information systems~Recommender systems}
\ccsdesc[300]{Human-centered computing~Web-based interaction}

\keywords{Information Foraging Theory, Information Retrieval, Recommendation System}

\maketitle

\section{Introduction}
Searching the Web is an important part of many people's everyday life. Retrieved online content is usually the outcome of generic user searches include textual documents, images, etc. As of now, the general searching method users employ is based on keywords, which is supported by almost every commercial search engine. Often users are also pointed to information by means of recommendation, which can for instance be based on similarity of documents or user profiles. To improve the effectiveness of search, there is an increasing interest in personalized or user-dependent search~\cite{dou2007large}. Personalized search systems expect to deduce user search preferences received from user feedback, which is crucial in web searches and image recommendation. People often find it very challenging when searching for images as in various situations they only know which images are relevant after they see them. Their cognitive abilities can understand an image when they see it in front, but their mind has confined ability to manifest a rich object like an image. This regular conscious consumption of information leads to the problem of information overload, for information that we are interested in is much harder to locate. People, in general, reflect an image based on the images seen before. Textual and visual representations in search engine result pages (SERPs) can be perceived in the context of a seminal state within Information Foraging Theory~\cite{chi2001using,pirolli1999information,pirolli2007information}. Information Foraging Theory postulates that users look at those information patches which have the strongest scent, where the scent strength is estimated by textual and visual cues from the information environment, contemplating the cue's relevance to the search task.
  %Articulating queries can be an art in an image search engine. 
After users start interacting with text-based recommender engines they provide the system with clues about their personalized preferences. The so gathered preferences are used to increase users' visual attention to enhance personalized image recommendation. The concept of foraging intervention, which we argue can be used in explainable recommendation, refers to a task of selecting the right item from a list of recommended documents or images, presenting that these interventions can coherently shape the information scent effects of user preferences, where the correct foraging strategy not only helps inferring those preferences, but also minimising users' cognitive load.

In this paper, we have collected real image data including visual bookmarks from a popular image-based social media network (Pinterest). In order to assess user attention within the recommended images from the test collection and the effects of making a choice among them, we investigated the impact of visual bookmarks pinned to every image by determining the information scent of images. Also, we explored the foraging effects of image recommendation in terms of user engagement and satisfaction. 

The contribution of this work is two-fold:
\begin{itemize}
    \item We propose a personalized recommendation system for image search that incorporates users' visual attention to recommended items; 
    \item We describe the user-dependent aspects we observe during foraging intervention across various effects of scent on a recommendation.
\end{itemize}

\section{Related Work}
The work in this paper rests on prior research in various areas, particularly Information Foraging Theory from behavioral psychology, image-based recommender systems, and image representation and content classification from machine learning.

\paragraph{Information Foraging Theory:} Information Foraging Theory~\cite{pirolli1999information} aims to model the information retrieval behavior which includes how information seekers navigate through information environments such as the web and help users finding their search strategies. Based on this theory, the user behavior to forage in the webpages (which are our information patches, see below) for specific information by trailing the information features (cues) on the Web is drawn by the patch's scent (information clues). In general, the foraging theory is based on the cost (time spent in search) and benefit (information consumption) assessments, and contemplation that people or animals recline toward rational strategies to maximize their information access or energy over an expanse of a given time. To adopt IFT for information seeking behavior which includes locating valuable pieces of information (document, image or other forms of data), seekers need to constantly evaluate cues from the online content spread over the Web. To this end, IFT follows three major concepts, which are:  (i)~\emph{Information Patch} designates a physical and conceptual space~\cite{gardenfors2004conceptual} of information which includes a webpage or an image divided into several regions where each region\footnote{generally rectangular, or could be of different shape based on the selected region of the object in an image} is made of pixels; (ii)~\emph{Information Scent} refers to the user's individual semantic compatibility to information objects and the preferred paths while navigating among/between patches via cue to estimate which nearest navigation path negotiates the probable value of distinct information object. 
Examples of information scent are visual or textual representations of the content i.e., text labels, tags, color or font. And (iii) \emph{Information Diet} refers to the combined set of information that has some perceived value to a searcher, who then emulates the set of information and neglects the rest~\cite{pirolli2007information}. Unfavorable information is emulated if a searcher pursues a generalized diet that comprises every genre of information confronted. A searcher will then spend much time searching if the information diet is overly idiosyncratic, that is, only some genres of information are available in the information diet.

Liu et al.~\cite{liu2011information} investigated an adaptive user interaction framework by applying IFT to demonstrate the effects of image search experience based on various user types realized during the quantitative analyses of three derived evaluations. Based on different user types for content-based image retrieval, Liu et al.~\cite{liu2010applying} demonstrated an IFT inspired user classification model to understand the users' interaction by functioning the model on several interaction features collected from the screen capture of various user task types on a content-based image retrieval system. They evaluated the classification model by performing qualitative data analysis and found that the six characteristics in the model are consistent with those interaction features which built a preliminary practice to study user interaction/behavior via IFT.

\paragraph{Personalized Image Recommendation:} The recent advancements in personalized image recommendation pave the way for various image recommender systems which include image-aware and image-unaware recommendation models, specifically on social network data. These two types of image recommender systems overlay various schemes introduced in ~\cite{he2016vbpr,chen2017attentive}, which efficaciously opt out images from a large collection of candidates that fit user's preference. The first type, image-aware recommender models, solely focuses on image representations and user modeling~\cite{he2016vbpr,chen2017attentive}, where representing images expressively and differentially has become one of the motives for image recommendations. However, He and McAuley~\cite{he2016vbpr} developed a model to exploit a pre-trained deep neural network, which supports the extraction of visual semantic embeddings from matrices containing images' pixels. This method lacks the image visual features in an immense fine-grained level because of treating an image as a whole single object. Recent work~\cite{chen2017attentive} proposed a multimedia recommendation model with an attention network side-by-side, which considers capturing image segments with comparative importance. In fact, this technique splits an image into equal-sized regions with the exclusion of semantic objects. We reckon that user preference to a definite image is supported by the inclusion of object semantics and exclusion of semantic objects leads to fallacy in image selection, descending the entire effectiveness of image recommendation.

On the other hand, the image-unaware recommendation models are entirely based on user modeling rather than considering the visual features of images. For instance, a user-item interaction without image information was described in~\cite{rendle2009bpr} which introduces a pairwise learning algorithm with implicit feedback. However, few techniques are developed to pilot behavior patterns or user profiles systematically to renovate the performance of recommendations~\cite{jing2014recommendation,jiang2014scalable,li2014personalized}. Past work~\cite{sang2012right} introduced a task (topic) sensitive model to characterise effects of the social network in a personalized image recommendation task. In our work we will investigate a content-based image recommendation in which we transform the image items in the representation space to recommend (or search) for identical items.

\section{Personalized Search Recommendation}
\begin{figure}[h!]
  \includegraphics[width=\linewidth]{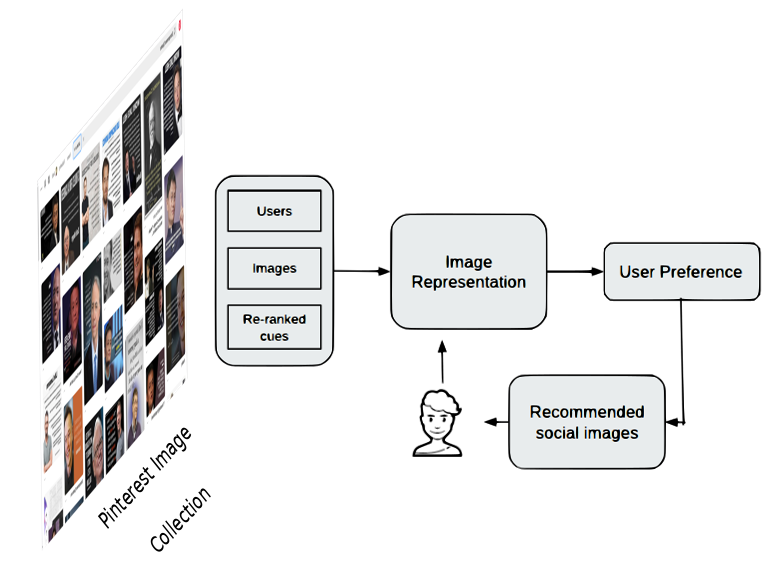}
  \caption{\label{fig:pir}\textbf{Personalized Image Recommendation}}
\end{figure}

We develop a personalized search recommendation engine for image searching based on what we call the \emph{User-Image-Cue model}. The schematic architecture is depicted as Figure~\ref{fig:pir}. In the User-Image-Cue model, users, images and (re-ranked) cues are framed on the left hand side of the figure within the interlinked graph. Their connection and role within the recommendation process is explained below.
The advantage of adopting content-based recommendation over collaborative filtering is that it does not have the cold start problem~\cite{lika2014facing} where a new item (or user) is introduced without previous history, as well as sufficient amount of data, whereas the latter exploits the users' correlation to make a recommendation. The content-based recommendation engine directs the image objects in the representation space, which permits to recommend for similar items.

\begin{figure}[h!]
  \includegraphics[width=\linewidth]{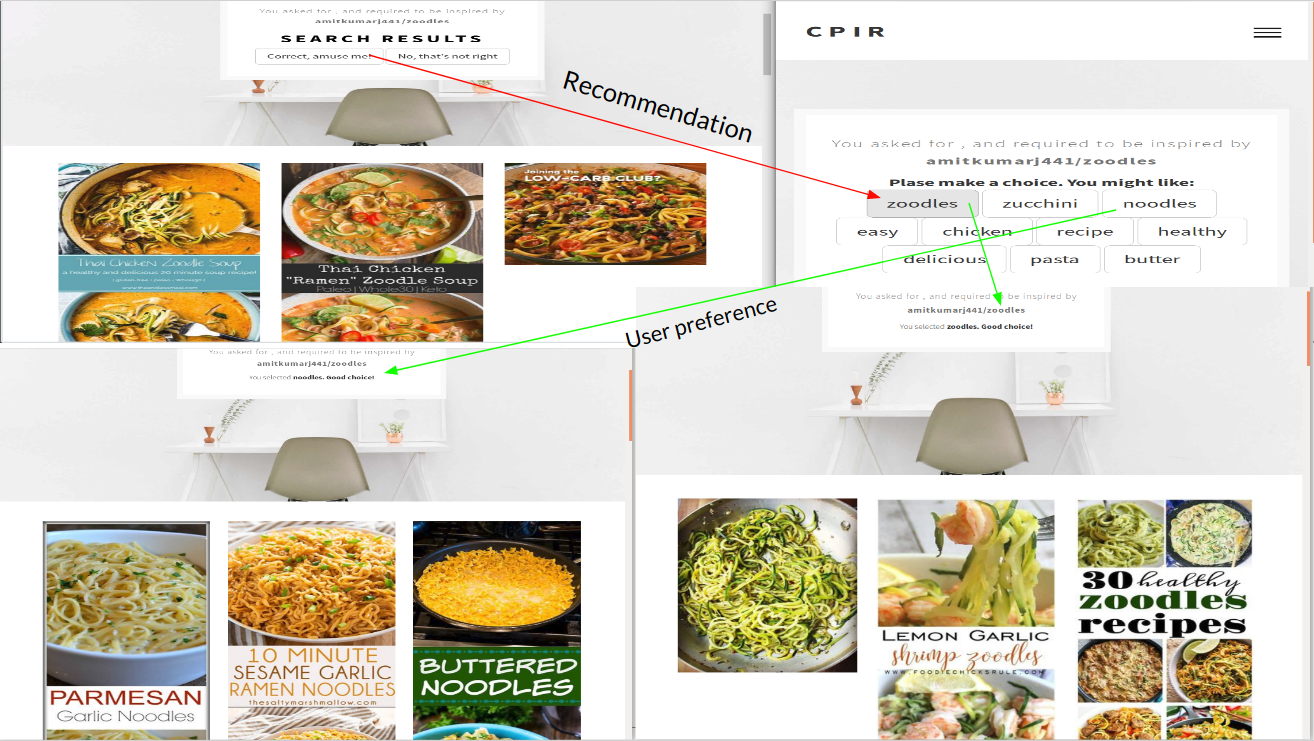}
  \caption{\label{fig:psri}\textbf{Personalized Search Recommendation Interface}}
\end{figure}

As per the above architecture, we make personalized search recommendations for image searching as shown in Figure~\ref{fig:psri} (which consists of four screen shots of our recommendation prototype). 

In the first phase (top left part of the figure), we have added a Pinterest board widget in the recommendation engine where the user inputs his/her board name\footnote{Similar to boards known in Pinterest, a board in our sense allows people to organize all their visual cues around diverse interests, ideas and plans} as a keyword-based query and it syncs the entire image collection in real-time from the specified Pinterest board. Users seeing the search result in the second phase (top right part of the figure) will be given several preferences based on their current search result to choose from, and if chosen the recommendation system again retrieves similar items (indicated by the green arrows in the figure pointing to different results in the bottom right and bottom left, respectively). Each and every image from the collection includes a cue associated with it.

\paragraph{Image Representation:} The image representation technique follows the hard-coded features of images which is a way to scale down computation, and as a simplified scenario of image embedding~\cite{frome2013devise}, similar to word embedding~(word2vec) where we first train the images with \{image, label\} dataset using a neural network, which then transform the image matrix representation (for instance 224x224x3) to a much smaller vector representation~(image2vec). This method can be used to compute the similarity between various images (or look for close vectors that depicts similar images). 

The motivation to adopt such interpretation is due to characterising the behavioral aspects of interactive elements (tags, cues and search interface, etc.) in recommendation system as opposed to focus on techniques developing a unified personalized RecSys, which is more or less based on learning or adopting features from user instead of bringing user-driven explainability in such system. Also, our proposed personalized recommendation engine can be supported with Pinterest image search algorithm~\cite{zhu2018demystifying} which has a quite better performance in terms of querying with text. 

\paragraph{Image Features:} We use different features to characterise images such as content, texture, color, and description/title. We train classifiers for each of these features on ImageNet, and employed each of them on images to extract the connected information.

We use a pre-trained ResNet50 model~\cite{he2016deep} to train our content classifier on ImageNet\footnote{https://pjreddie.com/darknet/imagenet/\#resnet50} which detects almost over 1000 different objects. Also, to predict the color (classification), we apply an unsupervised k-means clustering to match predominate colors to the generic color labels using html color scheme.

\section{Foraging Effects}
The first work on Visual Information Foraging can be traced in ~\cite{pirolli2001visual} to find information more quickly when there is a strong information scent~\cite{chi2001using} realised from cognitive perspectives. In this paper, we apply visual Information Foraging on a personalized image recommendation scenario to understand what drives a user to a generic search result (i.e., image) in terms of user engagement and satisfaction. 

We describe the effects of foraging in the context of image search which propagates implicit feedback for a content-based recommendation. In order to understand the personalized search recommendation interface by means of Information Foraging Theory, we formulate this recommendation system where the search engine result page (SERP) can be viewed as information patch together with all possible image views shapes a topology. The user aims to locate the interesting item in order to attain a decision in the foraging loop~\cite{pirolli2007information}. 

We hypothesize images as exemplary image patches\footnote{Images patches are image regions of a particular image when treated separately} that can be reached via cues while viewing an image content when it enables user cognitive beliefs. A user can activate such key beliefs via generating implicit cues to perceive ideas and plans for seeking, gathering and information consumption. In the same way animals believe on scents to forage~\cite{pirolli1999information} which is analogous to users following various kinds of cues in assessing image contents and navigating across patch spaces depend on images' scent. 

Images and tags form cues that correspond to an information scent. To acquire more information for locating the interesting image the cues compose the information diet and information access costs. The above discussion leads to three variables that can be interpreted via the personalized image recommendation system~-- the strength of the information scent, the effort involved in making conscious consumption about the image information and the information access cost for seeking extra information about an image. 

%Thus, we label those images which users' interested in as "1" and uninterested as "0", which is a binary prediction problem. 
Thus, from an IFT perspective, an image \emph{I} consists of $n$ image patches $I = \{I_{p_{i,1}}, I_{p_{i,2}},...,I_{p_{i,n}}\}$. For each image patch $I_{p_{i,j}} (j=1,2,...,n)$ we investigate those $m$ patches whose attention by the user $U$ is known and share the strong information scent with $I_{p{i,j}}$. Empirically, we compute the information scent of these image patches based on the frequency of user preferences for particular content. We use a psychometric scale such as Likert scale of 1-10 with '1' being least frequently to '10' being most frequently for evaluation of the recommendation system. The information scent of every user preference based on the recommendation is reported in Table~\ref{tab:scent}.

\section{Experimental Evaluation}
\subsection{Data Collection}
To evaluate the proposed recommendation system (RecSys), we compiled a real image dataset from Pinterest.com, a popular visual discovery sharing platform. We collected over 1116 images belonging to two categories of foods which includes \emph{Spaghetti Bolognese} and \emph{Zoodles}. We split the image data into 67\% train and 33\% test data. The associated information labels such as title and description with the images may indicate a very complex concept, where we use Naive Bayes to count the frequency of keywords (after data cleaning process).

\subsection{Results}
This section reports the evaluation result of user preferences based on the personalized image recommendation. We denote information scent and recommendation by ``IS'' and ``R'' respectively. In Table~\ref{tab:scent}, each recommendation ($R_1$, $R_2$, $R_3$, $R_4$, $R_5$) is ordered based on the strong information scent of user preferences, in which $R_i$ represents the inferred preferences of the user based on the $i$-th most liked images  (e.g., ``Bolognese'' and ``Zoodles'' for $R_1$) in the respective food categories collection. This means that those preferences with higher information scent are likely to be recommended to and attained by the searcher. This foraging-based observation makes users more likely to adopt visual bookmarks (visual cues) with little effort by hovering over recommended images instead of memorising the items themselves (with the latter discussed in~\cite{schnabel2016using}). This approach helps avoiding the searcher not to consume any sort of extra information diet (by memorising either items or buttons/tags).

If we interpret our scenario in terms of Information Foraging Theory, an image having either ``Bolognese'' or ``Zoodles'' (as in $R_1$) has a strong information scent. This means that such an image, presented as information patches, receives a comparatively large degree of attention by the user while likely to consume maximum information (information diet) and having lower information access costs, for instance in terms of the time spent on search.

%% Replaced the result table to free some spaces

%\begin{table}
%  \centering 
%   \begin{tabular}{|c|c|c|c|c|}
%    \hline 
%       \multirow{2}{*}{\textbf{Food Categories}} &
%       \multicolumn{2}{c|}{\textbf{Spaghetti Bolognese}} & 
%       \multicolumn{2}{c|}{\textbf{Zoodles}} \\
%      & User Preferences & IS & User Preferences & IS \\
%      \hline
%      R1 & Bolognese & 10 & Zoodles & 9 \\
%      \hline
%      R2 & Spaghetti & 7 & Zucchini & 8 \\
%      \hline
%      R3 & Recipe & 6 & Easy & 6 \\
%      \hline
%      R4 & Sauce & 6 & Pasta & 5 \\
%      \hline
%      R5 & Easy & 3 & Chicken & 5 \\
%      \hline
%   \end{tabular}
%   \caption{Information scent of User Preferences}
%\end{table}
\section{Conclusion \& Future Work}
This paper has investigated a personalized image recommendation system from an Information Foraging perspective. To this end, we conducted an empirical evaluation of user preferences in terms of information scent to get some first understanding of the effects of user attention during an image recommendation scenario in the context of IFT. This work found that:
\begin{enumerate}
    \item Information scent of an image has user-dependent aspects and users' scent of the same image can differ (For instance; "Bolognese" and "Spaghetti");
    \item The overall information scent of an image (as described in~\cite{loumakis2011image}) becomes stronger when adding cues;
    \item Reinforcing visual attention has a strong information scent, however, in some situations, the images' scent can exceed the cues' scent.
\end{enumerate}
%% TODO - Explainable recommendation by means of components from quantum-IR frameworks
We intend to scale up this study on a large test collection of images with varied categories that provide human expertise for characterising the images including its applicability in an explainable recommendation. Also, this work as the preliminary practice on applying IFT in recommendation system opens the door to evaluate such scenarios on other performance measures that can shed more light on efficiency and effectiveness by exploring interactions between information scent and cue strength.
\begin{table}
   \centering 
   \begin{tabular}{|c|c|c|c|c|}
    \hline 
       \multirow{2}{*}{\textbf{Food Categories}} &
       \multicolumn{2}{c|}{\textbf{Spaghetti Bolognese}} & 
       \multicolumn{2}{c|}{\textbf{Zoodles}} \\
      & User Preferences & IS & User Preferences & IS \\
      \hline
      $R_1$ & Bolognese & 10 & Zoodles & 9 \\
      \hline
      $R_2$ & Spaghetti & 7 & Zucchini & 8 \\
      \hline
      $R_3$ & Recipe & 6 & Easy & 6 \\
      \hline
      $R_4$ & Sauce & 6 & Pasta & 5 \\
      \hline
      $R_5$ & Easy & 3 & Chicken & 5 \\
      \hline
   \end{tabular}
   \caption{\label{tab:scent}Information scent of User Preferences}
\end{table}

\begin{acks}
This work was carried out in the context of Quantum Access and Retrieval Theory (QUARTZ) project, which has received funding from the European Union's Horizon 2020 research and innovation programme under the Marie Sklodowska-Curie grant agreement No. 721321.
\end{acks}

\bibliographystyle{ACM-Reference-Format}
\bibliography{bibdb}


\begin{thebibliography}{21}

%%% the Web address.

\ifx \showCODEN    \undefined \def \showCODEN     #1{\unskip}     \fi
\ifx \showDOI      \undefined \def \showDOI       #1{#1}\fi
\ifx \showISBNx    \undefined \def \showISBNx     #1{\unskip}     \fi
\ifx \showISBNxiii \undefined \def \showISBNxiii  #1{\unskip}     \fi
\ifx \showISSN     \undefined \def \showISSN      #1{\unskip}     \fi
\ifx \showLCCN     \undefined \def \showLCCN      #1{\unskip}     \fi
\ifx \shownote     \undefined \def \shownote      #1{#1}          \fi
\ifx \showarticletitle \undefined \def \showarticletitle #1{#1}   \fi
\ifx \showURL      \undefined \def \showURL       {\relax}        \fi
% invisible to TeX
\providecommand\bibfield[2]{#2}
\providecommand\bibinfo[2]{#2}
\providecommand\natexlab[1]{#1}
\providecommand\showeprint[2][]{arXiv:#2}

\bibitem[\protect\citeauthoryear{Chen, Zhang, He, Nie, Liu, and Chua}{Chen
  et~al\mbox{.}}{2017}]%
        {chen2017attentive}
\bibfield{author}{\bibinfo{person}{Jingyuan Chen}, \bibinfo{person}{Hanwang
  Zhang}, \bibinfo{person}{Xiangnan He}, \bibinfo{person}{Liqiang Nie},
  \bibinfo{person}{Wei Liu}, {and} \bibinfo{person}{Tat-Seng Chua}.}
  \bibinfo{year}{2017}\natexlab{}.
\newblock \showarticletitle{Attentive collaborative filtering: Multimedia
  recommendation with item-and component-level attention}. In
  \bibinfo{booktitle}{\emph{Proceedings of the 40th International ACM SIGIR
  conference on Research and Development in Information Retrieval}}. ACM,
  \bibinfo{pages}{335--344}.
\newblock


\bibitem[\protect\citeauthoryear{Chi, Pirolli, Chen, and Pitkow}{Chi
  et~al\mbox{.}}{2001}]%
        {chi2001using}
\bibfield{author}{\bibinfo{person}{Ed~H Chi}, \bibinfo{person}{Peter Pirolli},
  \bibinfo{person}{Kim Chen}, {and} \bibinfo{person}{James Pitkow}.}
  \bibinfo{year}{2001}\natexlab{}.
\newblock \showarticletitle{Using information scent to model user information
  needs and actions and the Web}. In \bibinfo{booktitle}{\emph{Proceedings of
  the SIGCHI conference on Human factors in computing systems}}. ACM,
  \bibinfo{pages}{490--497}.
\newblock


\bibitem[\protect\citeauthoryear{Dou, Song, and Wen}{Dou et~al\mbox{.}}{2007}]%
        {dou2007large}
\bibfield{author}{\bibinfo{person}{Zhicheng Dou}, \bibinfo{person}{Ruihua
  Song}, {and} \bibinfo{person}{Ji-Rong Wen}.} \bibinfo{year}{2007}\natexlab{}.
\newblock \showarticletitle{A large-scale evaluation and analysis of
  personalized search strategies}. In \bibinfo{booktitle}{\emph{Proceedings of
  the 16th international conference on World Wide Web}}. ACM,
  \bibinfo{pages}{581--590}.
\newblock


\bibitem[\protect\citeauthoryear{Frome, Corrado, Shlens, Bengio, Dean, Mikolov,
  et~al\mbox{.}}{Frome et~al\mbox{.}}{2013}]%
        {frome2013devise}
\bibfield{author}{\bibinfo{person}{Andrea Frome}, \bibinfo{person}{Greg~S
  Corrado}, \bibinfo{person}{Jon Shlens}, \bibinfo{person}{Samy Bengio},
  \bibinfo{person}{Jeff Dean}, \bibinfo{person}{Tomas Mikolov},
  {et~al\mbox{.}}} \bibinfo{year}{2013}\natexlab{}.
\newblock \showarticletitle{Devise: A deep visual-semantic embedding model}. In
  \bibinfo{booktitle}{\emph{Advances in neural information processing
  systems}}. \bibinfo{pages}{2121--2129}.
\newblock


\bibitem[\protect\citeauthoryear{Gardenfors}{Gardenfors}{2004}]%
        {gardenfors2004conceptual}
\bibfield{author}{\bibinfo{person}{Peter Gardenfors}.}
  \bibinfo{year}{2004}\natexlab{}.
\newblock \showarticletitle{Conceptual spaces as a framework for knowledge
  representation}.
\newblock \bibinfo{journal}{\emph{Mind and Matter}} \bibinfo{volume}{2},
  \bibinfo{number}{2} (\bibinfo{year}{2004}), \bibinfo{pages}{9--27}.
\newblock


\bibitem[\protect\citeauthoryear{He, Zhang, Ren, and Sun}{He
  et~al\mbox{.}}{2016}]%
        {he2016deep}
\bibfield{author}{\bibinfo{person}{Kaiming He}, \bibinfo{person}{Xiangyu
  Zhang}, \bibinfo{person}{Shaoqing Ren}, {and} \bibinfo{person}{Jian Sun}.}
  \bibinfo{year}{2016}\natexlab{}.
\newblock \showarticletitle{Deep residual learning for image recognition}. In
  \bibinfo{booktitle}{\emph{Proceedings of the IEEE conference on computer
  vision and pattern recognition}}. \bibinfo{pages}{770--778}.
\newblock


\bibitem[\protect\citeauthoryear{He and McAuley}{He and McAuley}{2016}]%
        {he2016vbpr}
\bibfield{author}{\bibinfo{person}{Ruining He} {and} \bibinfo{person}{Julian
  McAuley}.} \bibinfo{year}{2016}\natexlab{}.
\newblock \showarticletitle{VBPR: visual bayesian personalized ranking from
  implicit feedback}. In \bibinfo{booktitle}{\emph{Thirtieth AAAI Conference on
  Artificial Intelligence}}.
\newblock


\bibitem[\protect\citeauthoryear{Jiang, Cui, Wang, Zhu, and Yang}{Jiang
  et~al\mbox{.}}{2014}]%
        {jiang2014scalable}
\bibfield{author}{\bibinfo{person}{Meng Jiang}, \bibinfo{person}{Peng Cui},
  \bibinfo{person}{Fei Wang}, \bibinfo{person}{Wenwu Zhu}, {and}
  \bibinfo{person}{Shiqiang Yang}.} \bibinfo{year}{2014}\natexlab{}.
\newblock \showarticletitle{Scalable recommendation with social contextual
  information}.
\newblock \bibinfo{journal}{\emph{IEEE Transactions on Knowledge and Data
  Engineering}} \bibinfo{volume}{26}, \bibinfo{number}{11}
  (\bibinfo{year}{2014}), \bibinfo{pages}{2789--2802}.
\newblock


\bibitem[\protect\citeauthoryear{Jing, Zhang, Wu, Wang, Feng, and Wang}{Jing
  et~al\mbox{.}}{2014}]%
        {jing2014recommendation}
\bibfield{author}{\bibinfo{person}{Yuchen Jing}, \bibinfo{person}{Xiuzhen
  Zhang}, \bibinfo{person}{Lifang Wu}, \bibinfo{person}{Jinqiao Wang},
  \bibinfo{person}{Zemeng Feng}, {and} \bibinfo{person}{Dan Wang}.}
  \bibinfo{year}{2014}\natexlab{}.
\newblock \showarticletitle{Recommendation on Flickr by combining community
  user ratings and item importance}. In \bibinfo{booktitle}{\emph{2014 IEEE
  International Conference on Multimedia and Expo (ICME)}}. IEEE,
  \bibinfo{pages}{1--6}.
\newblock


\bibitem[\protect\citeauthoryear{Li, Luo, and Mei}{Li et~al\mbox{.}}{2014}]%
        {li2014personalized}
\bibfield{author}{\bibinfo{person}{Yuncheng Li}, \bibinfo{person}{Jiebo Luo},
  {and} \bibinfo{person}{Tao Mei}.} \bibinfo{year}{2014}\natexlab{}.
\newblock \showarticletitle{Personalized image recommendation for web search
  engine users}. In \bibinfo{booktitle}{\emph{2014 IEEE International
  Conference on Multimedia and Expo (ICME)}}. IEEE, \bibinfo{pages}{1--6}.
\newblock


\bibitem[\protect\citeauthoryear{Lika, Kolomvatsos, and Hadjiefthymiades}{Lika
  et~al\mbox{.}}{2014}]%
        {lika2014facing}
\bibfield{author}{\bibinfo{person}{Blerina Lika}, \bibinfo{person}{Kostas
  Kolomvatsos}, {and} \bibinfo{person}{Stathes Hadjiefthymiades}.}
  \bibinfo{year}{2014}\natexlab{}.
\newblock \showarticletitle{Facing the cold start problem in recommender
  systems}.
\newblock \bibinfo{journal}{\emph{Expert Systems with Applications}}
  \bibinfo{volume}{41}, \bibinfo{number}{4} (\bibinfo{year}{2014}),
  \bibinfo{pages}{2065--2073}.
\newblock


\bibitem[\protect\citeauthoryear{Liu, Mulholland, Song, Uren, and
  R{\"u}ger}{Liu et~al\mbox{.}}{2010}]%
        {liu2010applying}
\bibfield{author}{\bibinfo{person}{Haiming Liu}, \bibinfo{person}{Paul
  Mulholland}, \bibinfo{person}{Dawei Song}, \bibinfo{person}{Victoria Uren},
  {and} \bibinfo{person}{Stefan R{\"u}ger}.} \bibinfo{year}{2010}\natexlab{}.
\newblock \showarticletitle{Applying information foraging theory to understand
  user interaction with content-based image retrieval}. In
  \bibinfo{booktitle}{\emph{Proceedings of the third symposium on Information
  interaction in context}}. ACM, \bibinfo{pages}{135--144}.
\newblock


\bibitem[\protect\citeauthoryear{Liu, Mulholland, Song, Uren, and
  R{\"u}ger}{Liu et~al\mbox{.}}{2011}]%
        {liu2011information}
\bibfield{author}{\bibinfo{person}{Haiming Liu}, \bibinfo{person}{Paul
  Mulholland}, \bibinfo{person}{Dawei Song}, \bibinfo{person}{Victoria Uren},
  {and} \bibinfo{person}{Stefan R{\"u}ger}.} \bibinfo{year}{2011}\natexlab{}.
\newblock \showarticletitle{An information foraging theory based user study of
  an adaptive user interaction framework for content-based image retrieval}. In
  \bibinfo{booktitle}{\emph{International Conference on Multimedia Modeling}}.
  Springer, \bibinfo{pages}{241--251}.
\newblock


\bibitem[\protect\citeauthoryear{Loumakis, Stumpf, and Grayson}{Loumakis
  et~al\mbox{.}}{2011}]%
        {loumakis2011image}
\bibfield{author}{\bibinfo{person}{Faidon Loumakis}, \bibinfo{person}{Simone
  Stumpf}, {and} \bibinfo{person}{David Grayson}.}
  \bibinfo{year}{2011}\natexlab{}.
\newblock \showarticletitle{This image smells good: effects of image
  information scent in search engine results pages}. In
  \bibinfo{booktitle}{\emph{Proceedings of the 20th ACM international
  conference on Information and knowledge management}}. ACM,
  \bibinfo{pages}{475--484}.
\newblock


\bibitem[\protect\citeauthoryear{Pirolli}{Pirolli}{2007}]%
        {pirolli2007information}
\bibfield{author}{\bibinfo{person}{Peter Pirolli}.}
  \bibinfo{year}{2007}\natexlab{}.
\newblock \bibinfo{booktitle}{\emph{Information foraging theory: Adaptive
  interaction with information}}.
\newblock \bibinfo{publisher}{Oxford University Press}.
\newblock


\bibitem[\protect\citeauthoryear{Pirolli and Card}{Pirolli and Card}{1999}]%
        {pirolli1999information}
\bibfield{author}{\bibinfo{person}{Peter Pirolli} {and} \bibinfo{person}{Stuart
  Card}.} \bibinfo{year}{1999}\natexlab{}.
\newblock \showarticletitle{Information foraging.}
\newblock \bibinfo{journal}{\emph{Psychological review}} \bibinfo{volume}{106},
  \bibinfo{number}{4} (\bibinfo{year}{1999}), \bibinfo{pages}{643}.
\newblock


\bibitem[\protect\citeauthoryear{Pirolli, Card, and Van Der~Wege}{Pirolli
  et~al\mbox{.}}{2001}]%
        {pirolli2001visual}
\bibfield{author}{\bibinfo{person}{Peter Pirolli}, \bibinfo{person}{Stuart~K
  Card}, {and} \bibinfo{person}{Mija~M Van Der~Wege}.}
  \bibinfo{year}{2001}\natexlab{}.
\newblock \showarticletitle{Visual information foraging in a focus+ context
  visualization}. In \bibinfo{booktitle}{\emph{Proceedings of the SIGCHI
  conference on Human factors in computing systems}}. ACM,
  \bibinfo{pages}{506--513}.
\newblock


\bibitem[\protect\citeauthoryear{Rendle, Freudenthaler, Gantner, and
  Schmidt-Thieme}{Rendle et~al\mbox{.}}{2009}]%
        {rendle2009bpr}
\bibfield{author}{\bibinfo{person}{Steffen Rendle}, \bibinfo{person}{Christoph
  Freudenthaler}, \bibinfo{person}{Zeno Gantner}, {and} \bibinfo{person}{Lars
  Schmidt-Thieme}.} \bibinfo{year}{2009}\natexlab{}.
\newblock \showarticletitle{BPR: Bayesian personalized ranking from implicit
  feedback}. In \bibinfo{booktitle}{\emph{Proceedings of the twenty-fifth
  conference on uncertainty in artificial intelligence}}. AUAI Press,
  \bibinfo{pages}{452--461}.
\newblock


\bibitem[\protect\citeauthoryear{Sang and Xu}{Sang and Xu}{2012}]%
        {sang2012right}
\bibfield{author}{\bibinfo{person}{Jitao Sang} {and}
  \bibinfo{person}{Changsheng Xu}.} \bibinfo{year}{2012}\natexlab{}.
\newblock \showarticletitle{Right buddy makes the difference: An early
  exploration of social relation analysis in multimedia applications}. In
  \bibinfo{booktitle}{\emph{Proceedings of the 20th ACM international
  conference on Multimedia}}. ACM, \bibinfo{pages}{19--28}.
\newblock


\bibitem[\protect\citeauthoryear{Schnabel, Bennett, Dumais, and
  Joachims}{Schnabel et~al\mbox{.}}{2016}]%
        {schnabel2016using}
\bibfield{author}{\bibinfo{person}{Tobias Schnabel}, \bibinfo{person}{Paul~N
  Bennett}, \bibinfo{person}{Susan~T Dumais}, {and} \bibinfo{person}{Thorsten
  Joachims}.} \bibinfo{year}{2016}\natexlab{}.
\newblock \showarticletitle{Using shortlists to support decision making and
  improve recommender system performance}. In
  \bibinfo{booktitle}{\emph{Proceedings of the 25th International Conference on
  World Wide Web}}. International World Wide Web Conferences Steering
  Committee, \bibinfo{pages}{987--997}.
\newblock


\bibitem[\protect\citeauthoryear{Zhu}{Zhu}{2018}]%
        {zhu2018demystifying}
\bibfield{author}{\bibinfo{person}{Linhong Zhu}.}
  \bibinfo{year}{2018}\natexlab{}.
\newblock \showarticletitle{Demystifying Core Ranking in Pinterest Image
  Search}.
\newblock \bibinfo{journal}{\emph{arXiv preprint arXiv:1803.09799}}
  (\bibinfo{year}{2018}).
\newblock


\end{thebibliography}

\appendix



\end{document}